\documentclass[11pt]{article}

\usepackage{times}
\usepackage{latexsym}
\usepackage[T1]{fontenc}
\usepackage[utf8]{inputenc}

\usepackage[preprint]{acl}

\usepackage{microtype}
\usepackage{graphicx}
\usepackage{subcaption}
\usepackage{algorithm}
\usepackage{algorithmic}

\usepackage{booktabs}

\usepackage{adjustbox}
\usepackage{multirow}
\usepackage{listings}
\usepackage{xcolor}

\lstdefinestyle{python}{
    language=Python,
    basicstyle=\ttfamily\scriptsize,
    keywordstyle=\color{blue},
    stringstyle=\color{red},
    commentstyle=\color{gray},
    showstringspaces=false,
    breaklines=true,
    frame=single,
    numbers=left,
    numberstyle=\tiny\color{gray}
}
\usepackage{tcolorbox}
\tcbuselibrary{breakable}
\lstdefinelanguage{json}{
    basicstyle=\ttfamily\small,
    breaklines=true,
    frame=single
}

\usepackage{hyperref}  

\usepackage{amsmath}
\usepackage{amssymb}
\usepackage{mathtools}
\usepackage{amsthm}

\usepackage[capitalize,noabbrev]{cleveref}

\theoremstyle{plain}

\theoremstyle{definition}

\theoremstyle{remark}

\usepackage[textsize=tiny]{todonotes}

\title{SWE-Edit: Rethinking Code Editing for Efficient SWE-Agent}

\author{
  \textbf{Yikai Zhang\textsuperscript{1,2}\thanks{\ \ Work done during internship at Microsoft.}}, \quad
  \textbf{Jiaxin Pei\textsuperscript{3}}, \quad
  \textbf{Kenan Li\textsuperscript{1}}, \quad
  \textbf{Qirui Jin\textsuperscript{1}}, \quad
  \textbf{Maoquan Wang\textsuperscript{1}}, \quad
  \textbf{Jin Pan\textsuperscript{2}}, \quad
  \textbf{Yu Kang\textsuperscript{1}}, \\
  \textbf{Shengyu Fu\textsuperscript{1}}, \quad
  \textbf{Elsie Nallipogu\textsuperscript{1}}, \quad
  \textbf{Junjie Hu\textsuperscript{2}}, \quad
  \textbf{Yufan Huang\textsuperscript{1}}, \quad
  \textbf{Zijian Jin\textsuperscript{1}} \\[6pt]
  \textsuperscript{1}Microsoft, Redmond, Washington, United States \\
  \textsuperscript{2}Department of Computer Sciences, University of Wisconsin--Madison \\
  \textsuperscript{3}Stanford Institute for Human-Centered Artificial Intelligence (HAI), Stanford University \\[4pt]
  \texttt{zijianjin@microsoft.com} \quad
  \texttt{yufanhuang@microsoft.com} \quad
  \texttt{ykzhang@cs.wisc.edu}
}

\begin{document}
\maketitle

\begin{abstract}
Large language model agents have made strong progress on software engineering, yet current systems suffer from a context coupling problem: the standard code editing interface conflates code inspection, modification planning, and edit execution within a single context window, forcing agents to interleave exploratory viewing with strictly formatted edit generation. Irrelevant context accumulates and edit reliability degrades. We propose \textbf{SWE-Edit}, which decomposes the editing interface into two specialized subagents: a Viewer that extracts task-relevant code on demand, and an Editor that executes modifications from high-level natural language plans—letting the main agent focus on reasoning while delegating context-intensive operations to clean context windows. On SWE-Bench Verified, this decomposition raises resolve rate by 2.1\,pp and cuts inference cost by 17.9\%, with consistent gains across multiple reasoning-model families (Kimi-K2, MiniMax-M2.1, GLM-4.7). We further show that effective edit-format selection can be \emph{trained into a small model} rather than requiring frontier-scale capacity: GRPO training on Qwen3-8B with an adaptive find-replace/whole-file-rewrite policy improves edit success by 12.5\,pp and brings an 8B open-source editor to parity with GPT-5-nano on downstream SWE-Bench resolve rate. To enable rapid editor iteration, we release PR-Edit, a lightweight evaluation whose scores correlate strongly with SWE-Bench resolve rate. We release our code at \url{https://github.com/microsoft/SWE-Edit}.
\end{abstract}

\section{Introduction}
LLM-based coding agents can now solve real-world software engineering tasks by iteratively exploring codebases and refining solutions~\citep{yang2024swe, wang2024openhands}. Central to these systems is the \emph{code editing interface}~\citep{schluntz2025swebench}—the tools through which agents inspect files and apply modifications. However, current interfaces suffer from a fundamental \emph{context coupling problem}: they conflate code inspection, modification planning, and edit execution within a single context window, forcing agents to interleave exploratory viewing with strictly formatted edit generation.

This coupling creates a structural tension. For example, effective debugging requires broad exploration—viewing multiple files, tracing dependencies, testing hypotheses—yet each viewed file snippet persists in context regardless of its ultimate relevance. Meanwhile, generating correct code edits demands focused attention on precise locations and formats. Prior work establishes that LLM performance degrades when task-relevant information is buried within irrelevant context~\citep{shi2023large, liu2024lost}. For coding agents, exploration and precision are fundamentally at odds: a single agent cannot simultaneously optimize for comprehensive code understanding (which benefits from viewing many files) and reliable edit generation (which benefits from clean, focused context).

\begin{figure*}[t] 
    \vspace{0.1in}
    \centerline{\includegraphics[width=0.95\textwidth]{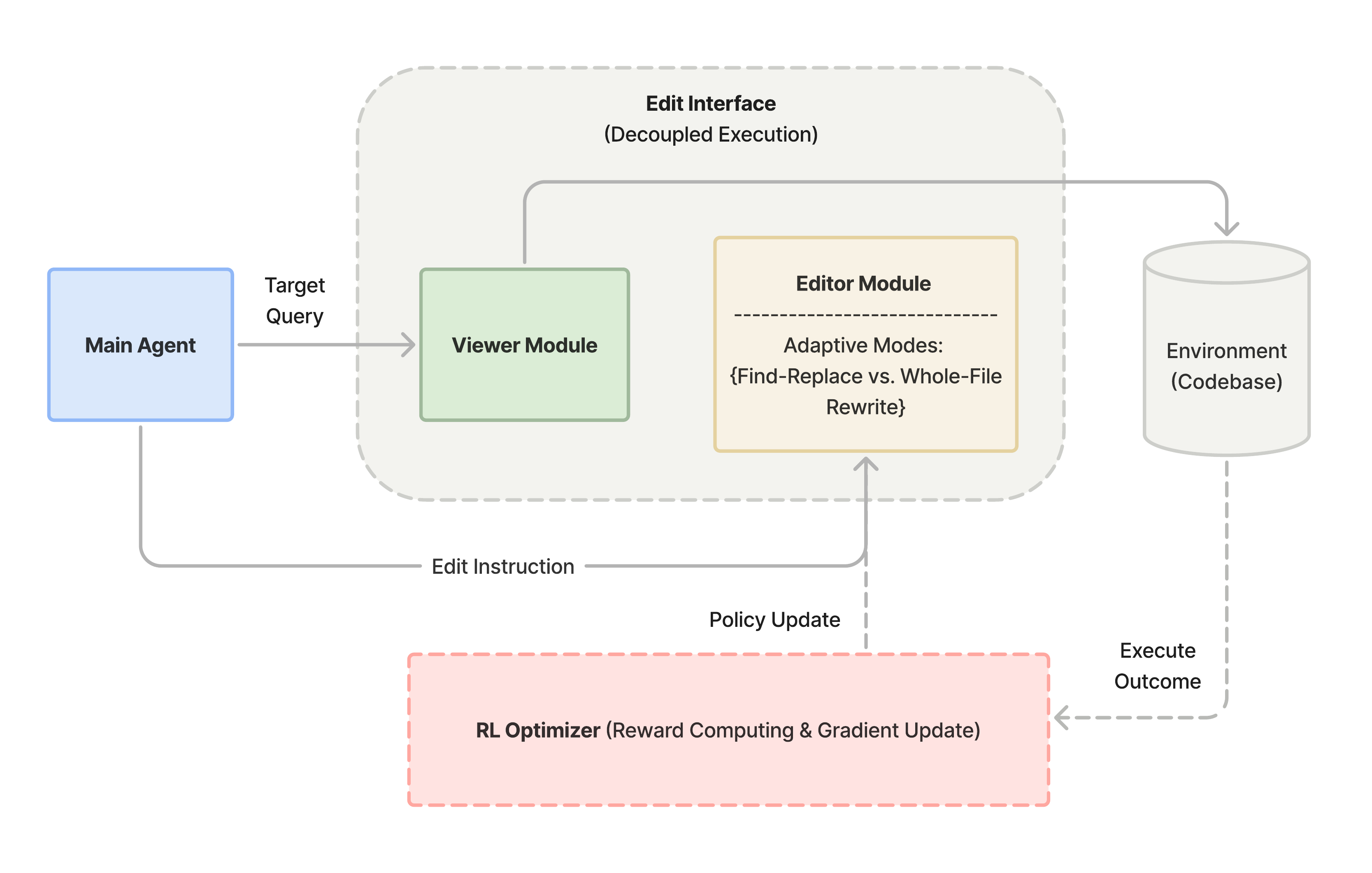}}
    \caption{Overview of the SWE-Edit framework. The main agent delegates inspection to the Viewer and modification to the Editor, each operating in a clean context. The Editor is additionally trained with GRPO to adaptively select between find-and-replace and whole-file rewrite.}
    \label{fig:swe_edit_architecture}
\end{figure*}

The code editing interface comprises two core operations that manifest this tension. The \textbf{view} operation allows agents to inspect file contents, but since agents cannot see code before viewing, they must explore incrementally—inevitably accumulating irrelevant context. The \textbf{edit} operation presents orthogonal challenges. The dominant \emph{find-replace} format requires exact string matching; a single whitespace mismatch causes the edit to fail. The alternative \emph{whole-file rewrite} avoids matching errors but incurs prohibitive token costs for long files. More fundamentally, reasoning about \emph{what} to modify and generating \emph{properly formatted} edit instructions are cognitively distinct capabilities: strong reasoning models like OpenAI o1~\citep{jaech2024openai} excel at describing solutions but frequently fail to produce correctly formatted edits~\citep{aider_architect}. Even frontier models like GPT-5\footnote{Throughout this paper, GPT-5 refers to the model with reasoning effort set to high.}~\citep{openai2025gpt5} exhibit non-trivial format-error rates on the Aider Polyglot code editing leaderboard~\citep{aider_polyglot}---a reliability gap largely overlooked when evaluating agents on end-to-end code editing tasks~\citep{aider_polyglot,jimenez2023swe}, where the cost of a single mis-formatted edit (a wasted reasoning trajectory) is hidden inside the resolve rate.

We propose \textbf{SWE-Edit} (Figure~\ref{fig:swe_edit_architecture}), built on a single principle: the code editing interface should separate \emph{what the agent sees}, \emph{what it decides}, and \emph{how edits are executed}.
The closest prior design, Aider Architect~\citep{aider_architect}, decouples reasoning from edit formatting but does not address context pollution from exploratory inspection, nor does it conduct experiments at repository scale. SWE-Edit instead introduces a three-way decomposition (Viewer + main agent + Editor) at the interface level, leaving the main agent unchanged and applying directly to closed-source models.

The \emph{Viewer} receives a file path and a natural-language query and returns only task-relevant code, eliminating context pollution. The \emph{Editor} receives a file path and a natural-language edit instruction and executes the modification directly, decoupling reasoning from format-sensitive generation. On SWE-Bench Verified, this decomposition raises resolve rate by 2.1\,pp and cuts inference cost by 17.9\%, with edit-success gains that transfer across three additional reasoning-model families (\S\ref{sec:scaffolding_results}).

The decomposition makes the editor an \emph{independently optimizable} role, which raises the question of how to optimize it. End-to-end SWE-Bench iteration is prohibitive (\textasciitilde\$200 per run), so we release \textbf{PR-Edit}, a lightweight editor-level evaluation whose scores rank-order the four editor models we test consistently with their downstream SWE-Bench resolve rate (Pearson $r{=}0.98$, $n{=}4$; Figure~\ref{fig:correlation}). Using PR-Edit as a fast signal, we find that the \emph{editing format itself} should be a per-task decision: find-and-replace is token-efficient for localized changes but brittle for restructuring, while whole-file rewrite is robust but costly. We formulate editing-mode selection as a learnable policy and train Qwen3-8B~\citep{yang2025qwen3} with GRPO~\citep{shao2024deepseekmath} to choose adaptively. The trained 8B editor matches GPT-5-nano on PR-Edit and downstream resolve rate, while scaling the editor from GPT-5-mini to GPT-5 yields only +0.4\,pp resolve rate at 5.8$\times$ the editor cost (Table~\ref{tab:editor_scaling})---evidence that format-level decision-making is a learnable skill, not a property of scale.

\section{Related Work}
\paragraph{LLM-Based Software Engineering} 
Large language models have advanced from code completion~\citep{chen2021evaluating,austin2021program,jain2024livecodebench,zhuo2024bigcodebench} to code editing~\citep{aider_polyglot} and repository-level software engineering~\citep{jimenez2023swe}. Early pipelines~\citep{orwall2024moatless,xia2024agentless} decompose tasks into localization, repair, and validation, while agentic systems~\citep{yang2024swe,wang2024openhands} equip LLMs with tools for iterative codebase interaction. Our work instead redesign the code editing interface itself.

\paragraph{Multi-Agent SWE Systems} 
Multi-agent frameworks like MetaGPT~\citep{hong2023metagpt} and ChatDev~\citep{qian2024chatdev} assign distinct development roles to communicating agents, and \citet{anthropic2025multiagent} distributes research queries into parallelizable subtasks. For SWE tasks specifically, HyperAgent~\citep{phan2024hyperagent} decomposes the workflow across Planner/Navigator/Code Editor/Executor roles, and BOAD~\citep{xu2025boad} automatically discovers orchestrator--subagent hierarchies for localization, editing, and validation via bandit optimization. These approaches decompose at the \emph{task} level---each subagent pursues a separable objective. SWE-Edit instead decomposes at the \emph{interface} level, isolating the cognitive interference between exploratory inspection and format-sensitive edit generation that arises within any monolithic editing tool. A complementary concern is context pollution, addressed via summarization-based context-folding~\citep{sun2025scaling} or callable context-management tools~\citep{liu2025context}. The closest prior design is Aider's \emph{Architect} mode~\citep{aider_architect}, which separates a reasoning model from a smaller editor; it addresses the reasoning--formatting bottleneck per edit but not the context pollution from prior exploration, and its high-capacity reasoning stage \emph{increases} cost. Our three-way decomposition (Viewer + main agent + Editor) addresses both bottlenecks and \emph{reduces} total cost (\S\ref{sec:scaffolding_results}).

\paragraph{Training and Evolving SWE-Agents} 
Parallel efforts improve SWE-agents through end-to-end training, adaptation, or skill-prior induction~\citep{xia2025live,sun2025scaling,wang2025swe,xie2025swe,yang2025kimi}. SWE-Edit differs in two ways: (i) our scaffolding operates at the \emph{interface} level, so Viewer and Editor replace standard file tools while keeping the main agent unchanged, applying directly to closed-source API models without retraining; and (ii) we train only the editor subagent and use the lightweight PR-Edit benchmark for iteration, rather than performing expensive end-to-end SWE-Bench RL.

\section{Method: The SWE-Edit Framework}
\label{sec:method}
\textbf{SWE-Edit} adopts a two-stage optimization framework to improve the efficiency and reliability of code editing agents. At the \emph{scaffolding level}, we decompose the code editing interface into specialized subagents, decoupling code inspection from modification to reduce context pollution, and decoupling high-level reasoning from format-sensitive generation to improve edit reliability (\S\ref{sec:scaffolding}). At the \emph{model level}, we train the editor to adaptively select between editing modes based on task characteristics, addressing the limitation that no single editing format performs optimally across modification types (\S\ref{sec:model_opt}).

\subsection{Scaffolding Optimization}
\label{sec:scaffolding}
The standard code editing interface couples two distinct operations: inspecting code to understand context, and reason about modifying code to actually implement changes. This coupling forces the main agent to accumulate exploratory context that may be irrelevant to the final edit. SWE-Edit restructures the code editing interface by decomposing it into two subagents—\textbf{Viewer} and \textbf{Editor}—each operating in a clean, specialized context.

\paragraph{Viewer Subagent}
The viewer receives a file path and a natural language query describing what information the agent seeks. Rather than returning raw file contents, it extracts and returns only the task-relevant code snippet. This filtering eliminates context pollution: the main agent receives precisely the information it needs without accumulating irrelevant code in its context window. To guard against over-pruning, the viewer prompt (Appendix~\ref{app:swe_edit_scaffold}) explicitly requires returning \emph{complete logical blocks} (Rule~3) with surrounding context (Rule~4). When the initial query is incomplete, the main agent can re-query the viewer with a refined request; in our SWE-Bench experiments, the agent issues an average of 7.49 viewer calls per instance, indicating natural recovery from incomplete retrievals.

\paragraph{Editor Subagent}
The editor receives a file path and a natural language edit instruction describing the desired modification. It executes the edit directly, without requiring the main agent to produce format-sensitive find-replace commands. This decouples high-level reasoning—deciding \emph{what} to change—from low-level generation—producing \emph{correctly formatted} edit syntax.

Both subagents are implemented using a smaller, cost-efficient model, while the main agent focuses purely on problem-solving and orchestration. Full implementation details and prompts are provided in Appendix~\ref{app:implementation}.

\subsection{Model Optimization}
\label{sec:model_opt}

Given the scaffolding decomposition, a natural question arises: what makes an effective editor? Prior systems commit statically to a single editing format—find-replace in Claude's str\_replace\_editor, Aider, and most agentic coding systems. Yet this is suboptimal: find-replace is token-efficient for localized changes but brittle to whitespace mismatches, while whole-file rewrite is robust to complex restructuring but costly and error-prone for long files.
\begin{figure*}[t] 
    \vspace{0.1in}
    \centerline{\includegraphics[width=0.95\textwidth]{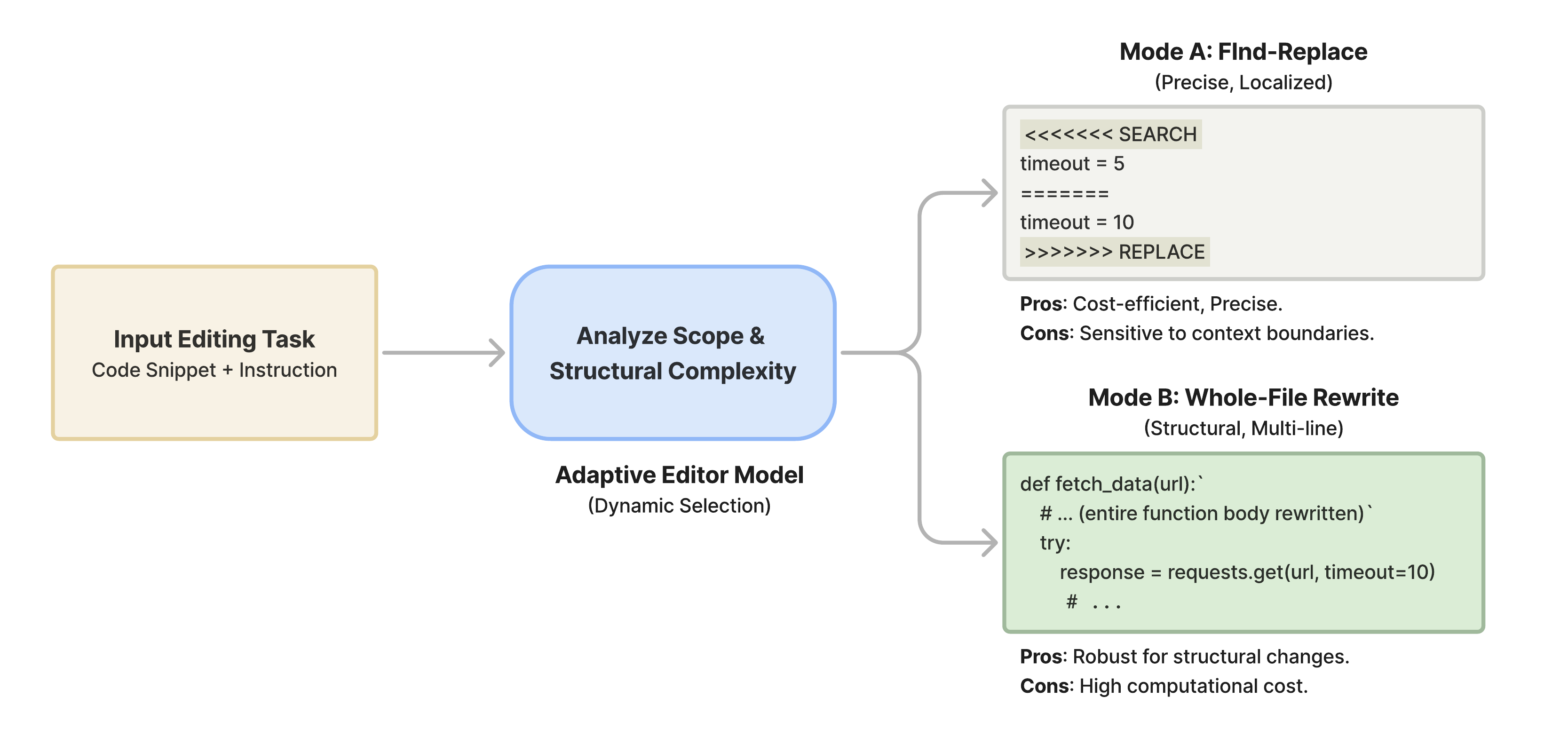}}
    \caption{Adaptive editing mode selection. The editor analyzes task characteristics to choose between find-replace (token-efficient but matching-sensitive) and whole-file rewrite (robust but costly), enabling optimal strategy selection based on edit scope and complexity.}
    \vspace{0.2in}
    \label{fig:adaptive_mechanism}
\end{figure*}

We observe that \textbf{the choice of editing format is itself a decision that should be conditioned on the task} (Figure~\ref{fig:adaptive_mechanism}): localized changes favor find-replace's token efficiency, while complex restructuring favors whole-file rewrite's robustness. To our knowledge, this has not been previously formulated as a learnable policy. We address this gap by framing editing mode selection as a single-step decision problem: given file contents $c$ and edit instruction $q$, the editor jointly selects mode $m \in \{\texttt{find-replace}, \texttt{whole-file-rewrite}\}$ and generates the corresponding output. We optimize this policy using GRPO~\citep{shao2024deepseekmath}, with a \emph{normalized match} reward that compares model output against ground truth after canonicalizing whitespace and removing comments (Listing~\ref{lst:normalize}). While this reward is an imperfect proxy for semantic correctness, we validate its alignment
empirically: models achieving higher normalized match scores also achieve higher GPT Grader accuracy on PR-Edit and, critically, higher resolve rates on SWE-Bench Verified where evaluation is fully execution-based (\S\ref{sec:model_results}).

\section{Experiments}
\label{sec:experiments}
Our experiments validate two core claims: (1) \textbf{interface decomposition} yields stable improvements in performance and cost efficiency over monolithic agent designs, and (2) \textbf{RL-based adaptive editing} learns robust mode selection policies, yielding more efficient editor models.

\subsection{General Setup}
\label{sec:setup}

\paragraph{Evaluation}
We evaluate primarily on SWE-Bench Verified~\citep{jimenez2023swe}, a curated benchmark of 500 real-world GitHub issues. All configurations are run 3 times to reduce variance. Full details of our SWE-Bench evaluation settings are provided in Appendix~\ref{app:implementation}.

\paragraph{Models}
For scaffolding experiments (\S\ref{sec:scaffolding_results}), we use GPT-5 as the main agent and GPT-5-mini for both subagents by default. For model-level experiments (\S\ref{sec:model_results}), we train Qwen3-8B~\citep{yang2025qwen3} with GRPO~\citep{shao2024deepseekmath}.

\paragraph{Metrics}
We report (1) \textbf{Resolve Rate}—percentage of issues where the agent's patch passes all tests; (2) \textbf{Total Cost}—inference cost across all components; (3) \textbf{Edit Success Rate}—percentage of edit operations without formatting errors; (4) \textbf{Viewer/Editor Calls}-average number of viewer/editor tool calls per instance. We additionally track a detailed breakdown of token usage and inference cost. The full experiment table is presented in Appendix~\ref{app:detailed_metrics}.


\subsection{Scaffolding-Level Results: Decomposition Yields Cost-Performance Synergy}
\label{sec:scaffolding_results}

\begin{table*}[t]
\caption{Main results on SWE-Bench Verified (500 instances, 3 runs averaged). The Baseline uses Anthropic's \texttt{str\_replace\_editor}; ``+ Viewer'' replaces only its \texttt{view} sub-command with the Viewer subagent (edits still use the baseline tool); ``+ Editor'' replaces only its \texttt{edit} sub-command with the Editor subagent (file inspection still uses the baseline tool); SWE-Edit replaces both via the unified \texttt{llm\_editor} (Appendix~\ref{app:swe_edit_scaffold}). Accordingly, ``Viewer Calls'' (resp.\ ``Editor Calls'') counts Viewer subagent invocations where present and \texttt{view} (resp.\ \texttt{edit}) sub-command invocations otherwise. SWE-Edit improves resolve rate (+2.1\,pp) and edit reliability (+3.5\,pp) while reducing cost by 17.9\%. Per-run standard deviations are shown in Table~\ref{tab:main_results_std} and confirm that the reported gains are stable across seeds (resolve rate $\pm 0.0$ for SWE-Edit).}
\label{tab:main_results}
\vspace{0.5em}
\begin{center}
\begin{tabular}{lccccc}
\toprule
\textbf{Configuration} & \textbf{Resolved (\%)} & \textbf{Cost (\$)} & \textbf{Viewer Calls} & \textbf{Editor Calls} & \textbf{Edit Succ. (\%)} \\
\midrule
Baseline & 69.9 & 243.7 & 5.78 & 2.86 & 93.4 \\
\quad + Viewer & 70.3 \small{\textcolor{teal}{(+0.4)}} & 225.0 \small{\textcolor{teal}{(-7.7\%)}} & 4.26 \small{\textcolor{teal}{(-1.52)}} & 2.75 \small{\textcolor{teal}{(-0.11)}} & 94.3 \small{\textcolor{teal}{(+0.9)}} \\
\quad + Editor & 71.3 \small{\textcolor{teal}{(+1.4)}} & 268.3 \small{\textcolor{red}{(+10.1\%)}} & 7.78 \small{\textcolor{red}{(+2.00)}} & 2.33 \small{\textcolor{teal}{(-0.53)}} & 96.1 \small{\textcolor{teal}{(+2.7)}} \\
\midrule
\textbf{SWE-Edit} & \textbf{72.0} \small{\textcolor{teal}{(+2.1)}} & \textbf{200.1} \small{\textcolor{teal}{(-17.9\%)}} & 7.49 \small{\textcolor{red}{(+1.71)}} & 2.37 \small{\textcolor{teal}{(-0.49)}} & \textbf{96.9} \small{\textcolor{teal}{(+3.5)}} \\
\bottomrule
\end{tabular}
\end{center}
\vspace{-1em}
\end{table*}

A natural concern with subagent decomposition is the overhead of additional inference calls. We show that the viewer and editor subagents provide complementary benefits that combine synergistically: SWE-Edit achieves \emph{both} higher resolve rate and lower cost than the monolithic baseline.\par
\vspace{0.5ex} 
\noindent \textbf{Viewer: Reducing Context Pollution}
As shown in Table~\ref{tab:main_results}, adding the viewer reduces total cost by 7.7\% (\$243.7 $\rightarrow$ \$225.0) while slightly improving resolve rate (+0.4\,pp). Two mechanisms drive this reduction: (1) the viewer uses a smaller model (GPT-5-mini) to process full file contents, and (2) by extracting only task-relevant snippets, it reduces context pollution in the main agent's window. The viewer returns on average only 39.7\% of the requested content---a 60.3\% reduction in code surface---lowering the main agent's non-cached input from 276.7K to 237.1K tokens ($-14.3\%$; Table~\ref{tab:detailed_results_appendix}). When combined with the editor (see \emph{Synergistic Effect} below), this reduction compounds to 34.5\% (276.7K $\rightarrow$ 181.3K), as the editor further eliminates costly formatting retries. Viewer calls themselves decrease from 5.78 to 4.26 per instance, as more focused responses eliminate redundant file re-reads.
Controlled comparisons against dense retrieval and BM25 confirm that the LLM viewer substantially outperforms classical retrieval (Appendix~\ref{app:viewer_retrieval}), and replacing GPT-5-mini with GPT-5 yields identical performance, confirming a smaller model suffices.

\noindent \textbf{Editor: Enhancing Edit Precision and Reliability}
The editor subagent in isolation improves resolve rate from 69.9\% to 71.3\% (\textcolor{teal}{+1.4\%}) and edit success rate from 93.4\% to 96.1\% (\textcolor{teal}{+2.7\%}), confirming that decoupling reasoning from syntactic execution resolves formatting errors and facilitates logically sound patches. Editor calls decrease from 2.86 to 2.33 per instance---higher reliability means fewer retries. However, this comes at increased cost (\textcolor{red}{+10.1\%}): the main agent becomes more exploratory when delegating edits (viewer calls rise from 5.78 to 7.78), suggesting that natural language edit instructions encourage gathering more context before committing to a plan. This behavioral shift further motivates combining the editor with the viewer subagent.

\begin{figure}[ht]
\begin{center}
    \centerline{\includegraphics[width=\columnwidth]{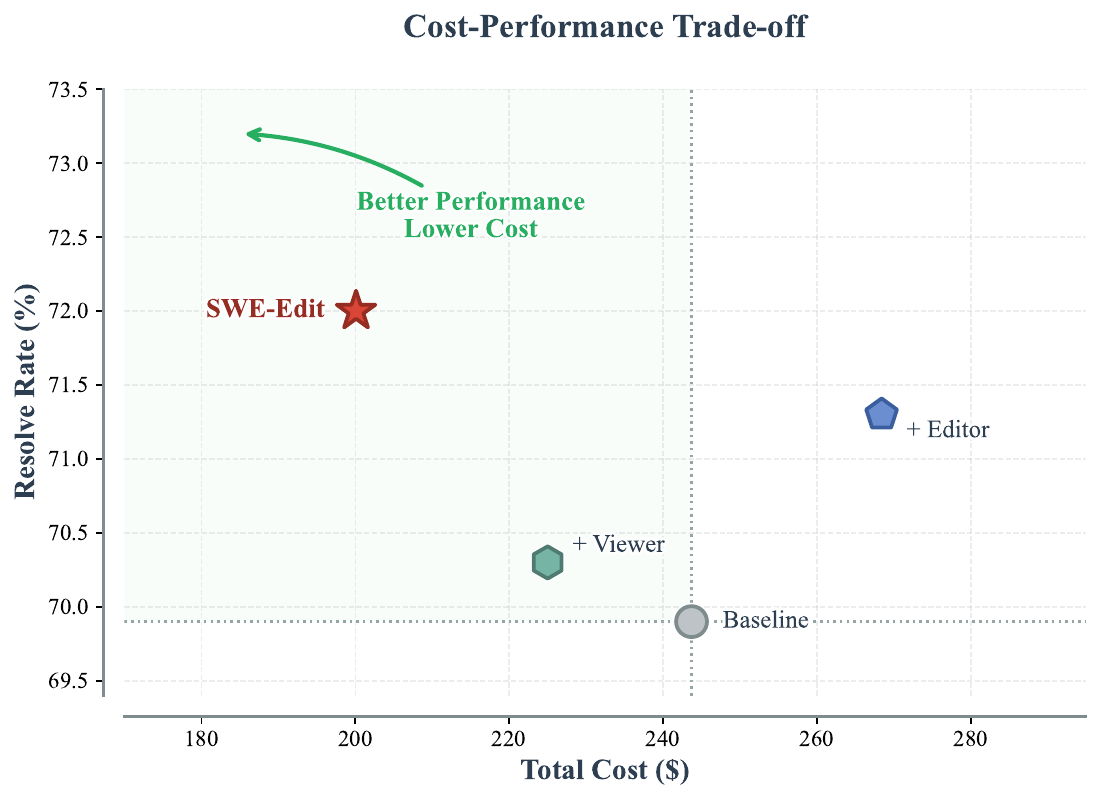}}
    \caption{Cost-performance trade-off on SWE-Bench Verified. Dashed lines indicate baseline performance. The viewer reduces cost (leftward), the editor improves resolve rate (upward), and SWE-Edit achieves both, occupying the high-performance, low-cost quadrant.}
    \label{fig:main_results}
\end{center}
\end{figure}

\paragraph{Synergistic Effect: Breaking the Accuracy-Cost Trade-off}
The complete SWE-Edit framework combines both subagents, achieving the highest resolve rate (72.0\%) at the lowest cost (\$200.1)---improving over baseline by \textcolor{teal}{2.1\,pp} absolute while reducing cost by \textcolor{teal}{17.9\%} (relative). Edit success rate reaches 96.9\% with only 2.37 editor calls per instance. Although viewer calls remain elevated (7.49 vs.\ 5.78 baseline), the viewer processes these with a smaller model and returns filtered snippets, yielding net savings. The two subagents address \emph{complementary} cost drivers, and their interaction is non-additive in a useful direction. Introducing the Editor alone makes the main agent more exploratory (file-inspection calls rise from 5.78 to 7.78 in the ``+ Editor'' row of Table~\ref{tab:main_results}), which would otherwise inflate context cost. Adding the Viewer absorbs exactly this exploration burst by filtering each inspection through a smaller model. The net effect is that non-cached main-agent input tokens drop from 276.7K (Baseline) to 181.3K (SWE-Edit), a $34.5\%$ reduction (Table~\ref{tab:detailed_results_appendix}), while the Editor independently eliminates costly formatting retries. The combination shifts the Pareto frontier rather than trading one metric for another (Figure~\ref{fig:main_results}).

\paragraph{Generalization to Diverse Reasoning Models}
To verify that SWE-Edit's benefits extend beyond GPT-5, we evaluate on three recent reasoning models: Kimi-K2~\citep{moonshot2025kimik2thinking}, MiniMax-M2.1~\citep{minimax2025m21}, and GLM-4.7~\citep{zhipuai2025glm47}. Across all configurations, only the main-agent model is changed; the Viewer and Editor subagents are kept fixed at GPT-5-mini. Due to computational constraints, we run each configuration twice on the first 100 instances of SWE-Bench Verified (details in Appendix~\ref{app:opensource}).

\begin{table}[h]
\caption{Generalization to diverse reasoning models (100 instances, 2 runs). SWE-Edit consistently improves resolve rate and edit success across all models.}
\label{tab:opensource}
\setlength{\tabcolsep}{3pt}
\begin{center}
\begin{footnotesize}
\begin{tabular}{@{}llcc@{}}
\toprule
\textbf{Model} & \textbf{Config} & \textbf{Res.\,(\%)} & \textbf{Edit Succ.\,(\%)} \\
\midrule
Kimi K2 & Baseline & 56.7 & 75.6 \\
 & SWE-Edit & \textbf{59.4} {\scriptsize\textcolor{teal}{(+2.7)}} & \textbf{93.9} {\scriptsize\textcolor{teal}{(+18.3)}} \\
\midrule
MiniMax-M2.1 & Baseline & 58.8 & 82.0 \\
 & SWE-Edit & \textbf{62.9} {\scriptsize\textcolor{teal}{(+4.1)}} & \textbf{94.8} {\scriptsize\textcolor{teal}{(+12.8)}} \\
\midrule
GLM-4.7 & Baseline & 63.3 & 79.6 \\
 & SWE-Edit & \textbf{64.9} {\scriptsize\textcolor{teal}{(+1.6)}} & \textbf{95.9} {\scriptsize\textcolor{teal}{(+16.3)}} \\
\bottomrule
\end{tabular}
\end{footnotesize}
\end{center}
\vspace{-1em}
\end{table}

\begin{table*}[th]
\caption{Results on PR-Edit Benchmark. GRPO training substantially improves Qwen3-8B, achieving performance comparable to GPT-5-nano.}
\label{tab:pr_benchmark}
\vspace{0.5em}
\begin{center}
\begin{small}
\begin{tabular}{lccc}
\toprule
\textbf{Model} & \textbf{Format (\%)} & \textbf{GPT Grader (\%)} & \textbf{Norm. Match (\%)} \\
\midrule
Qwen3-8B & 76.8 & 56.0 & 32.0 \\
Qwen3-8B + GRPO & \textbf{90.4} & \textbf{68.4} & \textbf{38.8} \\
\midrule
GPT-5-nano & 89.8 & 66.4 & 38.8 \\
GPT-5-mini & 96.1 & 77.5 & 41.7 \\
GPT-5 & 98.1 & 77.2 & 44.1 \\
\bottomrule
\end{tabular}
\end{small}
\end{center}
\end{table*}

\begin{table*}[t]
\caption{Downstream performance on SWE-Bench Verified with different editor models. Higher PR-Edit scores predict better resolve rate, higher edit success, and lower main agent cost.}
\label{tab:downstream}
\vspace{0.5em}
\begin{center}
\begin{small}
\begin{tabular}{lcccc}
\toprule
\textbf{Editor Model} & \textbf{PR-Edit (\%)} & \textbf{Resolved (\%)} & \textbf{Agent Cost (\$)} & \textbf{Edit Succ. (\%)} \\
\midrule
Qwen3-8B & 56.0 & 68.5 & 231.7 & 68.6 \\
Qwen3-8B + GRPO & 68.4 & 69.9 & 215.9 & 81.1 \\
\midrule
GPT-5-nano & 66.4 & 70.0 & 207.1 & 82.0 \\
GPT-5-mini & 77.5 & 72.0 & 179.6 & 96.9 \\
\bottomrule
\end{tabular}
\end{small}
\end{center}
\vspace{-1em}
\end{table*}

As shown in Table~\ref{tab:opensource}, SWE-Edit consistently outperforms the baseline across all three models. The most striking improvement is in edit success rate: baseline configurations exhibit variable reliability (75.6\%--82.0\%), while SWE-Edit stabilizes at 93.9\%--95.9\% (+12.8 to +18.3\,pp). This confirms that the editor subagent's reliability benefits are model-agnostic---reasoning models, which often struggle more with strict formatting requirements, benefit even more from the decoupled architecture. Resolve rate improvements of +1.6 to +4.1\,pp demonstrate that the scaffolding-level gains transfer across model families; the smaller magnitude relative to edit-success gains reflects that format errors are one of several failure modes on the critical path to a passing patch.


\subsection{Model-Level Results: Training Adaptive Editors}
\label{sec:model_results}
Having established SWE-Edit's scaffolding benefits, we now turn to optimizing the editor itself. Under the proposed decomposition, a natural question arises: \textbf{How can we effectively train and select models to excel in the specialized role of an editor?}
We leverage the modularity of our framework to perform targeted RL on a Qwen3-8B backbone, transforming it into an adaptive editing subagent.

\subsubsection{Training Setup}

\paragraph{Data}
We curate training data from open-source GitHub pull requests across diverse repositories. For each PR, we extract the file content before and after merging, along with the git diff and PR message. We prompt GPT-4.1~\citep{openai2024gpt41} to generate natural language edit instructions conditioned on these artifacts, yielding 3.5K examples split into 2.8K training, 200 validation, and 500 held-out test instances. The split is performed at the pull-request level. To prevent leakage between training and downstream evaluation, we additionally verify that there is no repository overlap with SWE-Bench Verified and zero exact-code overlap between train and test files.

\paragraph{RL Training}
We fine-tune Qwen3-8B~\citep{yang2025qwen3} using GRPO~\citep{shao2024deepseekmath} on Slime~\citep{slime_github}. The model learns to select an editing mode and generate the corresponding output. Training runs for 520 rollout steps; each step samples 32 instances with 8 candidate outputs per instance. We use the normalized match reward described in \S\ref{sec:model_opt}.

\paragraph{Intermediate Evaluation (PR-Edit Benchmark)}
We reserve the 500 held-out examples as the \textbf{PR-Edit Benchmark}, a lightweight and efficient intermediate evaluation for editor models.
Compared to end-to-end evaluation on SWE-Bench Verified, PR-Edit is substantially cheaper and faster to run, enabling rapid iteration on editor design.
We report three metrics: (1) \emph{Format Success}; (2) \emph{GPT Grader} (GPT-4.1 assesses edit correctness); and (3) \emph{Normalized Match} (exact match after canonicalization). The GPT Grader is used only for PR-Edit; our main SWE-Bench evaluation is fully execution-based. As shown in Figure~\ref{fig:correlation}, PR-Edit scores predict the ordering of editor models on SWE-Bench (Pearson $r{=}0.98$ across the four editor models in Table~\ref{tab:downstream}). Although four points is a small sample for a correlation estimate, the ranking is strictly monotonic across all three metric pairs (PR-Edit vs.\ resolve rate, edit success, and main-agent cost), and we report this primarily as a model-selection signal rather than a precise correlation estimate.

\subsubsection{Evaluation of Adaptive Editor Training}
\label{sec:adaptive_editor_eval}
Table~\ref{tab:pr_benchmark} shows that GRPO training substantially improves Qwen3-8B: format success rises from 76.8\% to 90.4\% (\textcolor{teal}{+13.6\,pp}) and GPT Grader accuracy from 56.0\% to 68.4\% (\textcolor{teal}{+12.4\,pp}), exceeding GPT-5-nano across all metrics.

\begin{figure}[t]
\centering
\includegraphics[width=\columnwidth]{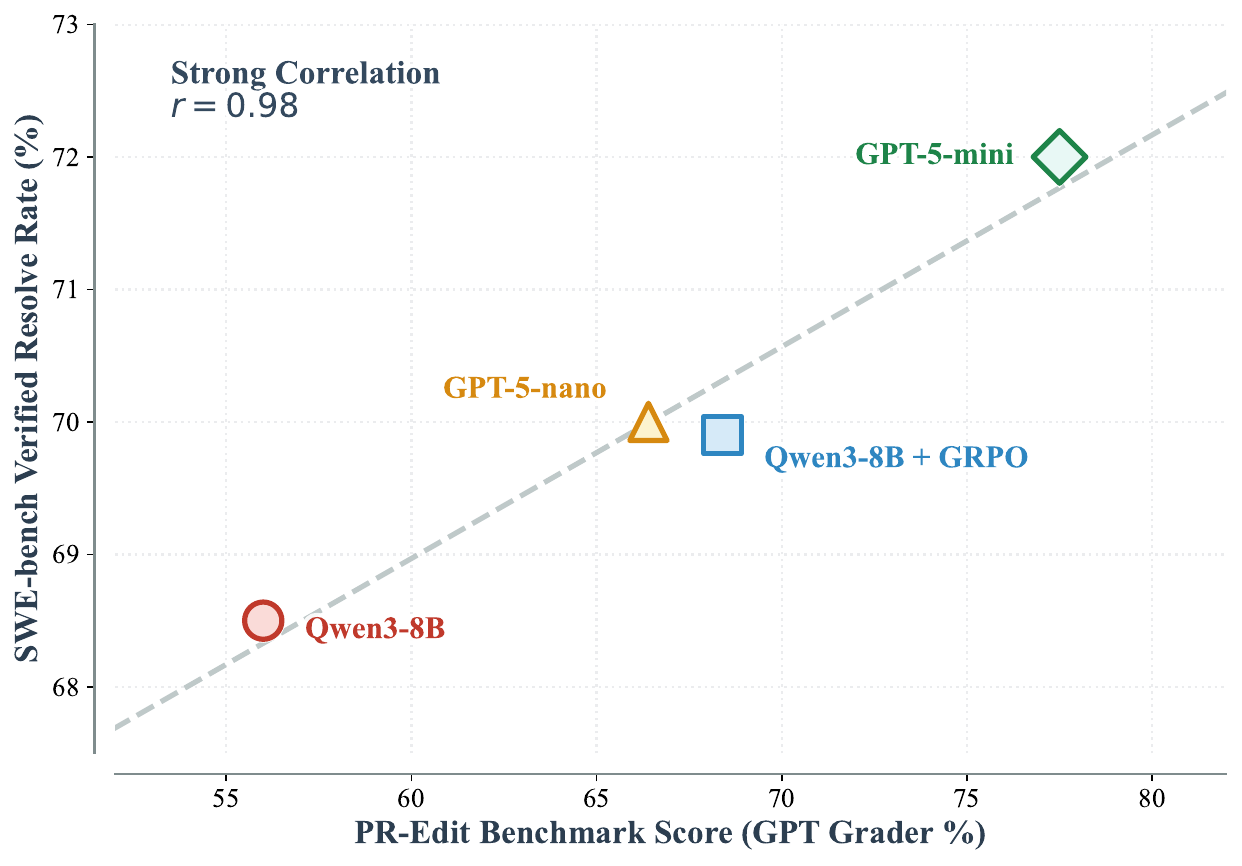}
\caption{PR-Edit benchmark scores correlate with downstream agent performance, enabling efficient editor model selection without full SWE-Bench evaluation.}
\label{fig:correlation}
\end{figure}

Critically, these editor-level gains translate to downstream performance.
As shown in Table~\ref{tab:downstream}, improvements on the PR-Edit Benchmark consistently correspond to stronger downstream performance, including higher resolve rates, higher edit success rates, and lower main-agent inference cost.
The GRPO-trained Qwen3-8B improves SWE-Bench resolve rate from 68.5\% to 69.9\% (\textcolor{teal}{+1.4\,pp}) and edit success from 68.6\% to 81.1\% (\textcolor{teal}{+12.5\,pp}), while reducing main-agent cost by 6.8\% (relative). Figure~\ref{fig:correlation} visualizes this correlation, confirming that PR-Edit serves as a reliable intermediate signal for editor quality.

\subsection{Ablation Studies}
\label{sec:ablations}

\paragraph{Adaptive vs.\ Fixed Edit Format}
We compare adaptive format selection against a fixed find-replace baseline---the dominant format in practice, used by Claude's \texttt{str\_replace\_editor}~\citep{schluntz2025swebench}, Aider~\citep{aider_polyglot}, and most agentic coding systems. We treat \emph{find-replace} and \emph{diff-patch} as variants of the same class of exact-match editing; the substantive design choice is between exact-match editing and whole-file rewrite. Table~\ref{tab:ablation_format} shows that adaptive selection outperforms the fixed baseline on both PR-Edit and downstream SWE-Bench evaluation, as localized changes are handled efficiently by find-replace while complex refactoring benefits from whole-file rewrite.

\begin{table}[t]
\caption{Ablation on edit format strategy. Adaptive selection outperforms fixed find-replace by learning to match strategy to task complexity. All metrics reported in percentages except Cost (in USD).}
\label{tab:ablation_format}
\vspace{0.5em}
\setlength{\tabcolsep}{4pt} 
\begin{center}
\begin{footnotesize}
\begin{tabular}{lcccc}
\toprule
\textbf{Strategy} & \textbf{PR-Edit} & \textbf{Resolved} & \textbf{Cost} & \textbf{Edit Succ.} \\
\midrule
Find-Replace & 67.0 & 69.4 & 244.7 & 80.2 \\
Adaptive (Ours) & \textbf{68.4} & \textbf{69.9} & \textbf{215.9} & \textbf{81.1} \\
\bottomrule
\end{tabular}
\end{footnotesize}
\end{center}
\vspace{-0.1in}
\end{table}

Training dynamics (Appendix~\ref{app:training_dynamics}, Figure~\ref{fig:training_dynamics}) reveal that the fixed find-replace policy starts with higher validation reward but the adaptive policy surpasses it, confirming that learning to invoke whole-file rewrite for the subset of tasks where find-replace struggles yields consistent gains.

\paragraph{Scaling the Viewer and Editor Model}
Replacing GPT-5-mini with GPT-5 in the viewer yields identical
performance, confirming a smaller model suffices
(Appendix~\ref{app:viewer_scaling}).

Replacing GPT-5-mini with GPT-5 as the editor improves resolve rate by only 0.4\% at 5.8$\times$ the cost (Appendix~\ref{app:editor_scaling}, Table~\ref{tab:editor_scaling}). In contrast, GRPO training yields a 12.5pp increase in edit success on the much smaller Qwen3-8B, indicating that format-level decision-making is a learnable skill rather than a capability that emerges only at scale.

\section{Conclusion}
We present SWE-Edit, a two-stage optimization framework to enhance the code editing interface in software engineering agents. Our three contributions form a self-contained pipeline. The scaffold creates a modular editor role with a clean interface, which improves agent performance on SWE tasks while saving significant inference cost. Once that role is isolated, the natural follow-up question is how to improve the model that fills it; doing so via end-to-end SWE-Bench iteration is prohibitively expensive (\textasciitilde\$200 and several hours per run), which motivates PR-Edit as a lightweight intermediate benchmark that correlates strongly with downstream agent performance (Figure~\ref{fig:correlation}). With PR-Edit in place, RL-based editor training becomes practical, and we show that GRPO with a normalized match reward yields a 12.5\,pp improvement in edit success on a small open-source backbone (Qwen3-8B), substantially exceeding what model scaling alone provides. Together, these three pieces—scaffold, benchmark, and training—provide a deployable and cost-efficient recipe for code-editing subagents.

\section*{Limitations}
One limitation of our current approach is that we train the editor model in isolation using a static reward signal derived from ground-truth edits. A natural extension is to train the editor within an end-to-end agentic reinforcement learning loop, where it receives feedback from the main agent's downstream success or failure. This would allow the editor to learn not just formatting correctness, but also properties that facilitate effective agent-editor collaboration—such as producing edits that are easier for the main agent to verify or debug when errors occur. A complementary direction is to replace the GPT Grader with an \emph{agent-as-a-judge} evaluator that assesses edits using broader repository context, rather than comparing two diffs in isolation.

\bibliography{example_paper}

\appendix

\section{Implementation Details}
\label{app:implementation}

This appendix provides full details of the agent scaffolding, tool definitions, and prompts used in our experiments.

\subsection{Baseline Agent Scaffolding}
\label{app:baseline_scaffold}

We adopt the reference agent scaffolding from Anthropic~\citep{schluntz2025swebench}, which equips the agent with two tools: \texttt{execute\_bash} for shell command execution and \texttt{str\_replace\_editor} for file operations. The editor tool provides sub-commands for viewing, creating, and editing files via exact string replacements. This reflects current best practices for agentic software engineering.

\paragraph{Tool Definitions.}
The baseline agent uses the following tools (Figure~\ref{lst:bash_tool}).

\begin{figure*}[t]
\begin{lstlisting}[language=json, caption={Bash tool schema.}, label={lst:bash_tool}, basicstyle=\ttfamily\scriptsize, breaklines=true, breakatwhitespace=false, postbreak=\mbox{\textcolor{red}{$\hookrightarrow$}\space}]
{
  "type": "function",
  "name": "execute_bash",
  "description": "Run commands in a bash shell\n* When invoking this tool, the contents of the \"command\" parameter does NOT need to be XML-escaped.\n* You don't have access to the internet via this tool.\n* You do have access to a mirror of common linux and python packages via apt and pip.\n\n### Command Execution\n* **Non-persistent**: Each shell tool call is executed in a fresh environment. Shell variables, working directory changes, and history are NOT preserved between calls.\n* **Timeout**: Commands have a default timeout of 120 seconds (max 300). Set the `timeout` parameter for long-running commands.\n* **One command at a time**: Chain multiple commands using `&&` (conditional), `;` (sequential), or `||` (on failure).\n\n### Long-running Commands\n* For commands that may run indefinitely (e.g., servers), run in background: `python3 app.py > server.log 2>&1 &`\n* For potentially long commands (installations, tests), set an appropriate `timeout` value.\n\n### Best Practices\n* **Avoid large outputs**: Commands producing massive output may be truncated.\n* **Directory verification**: Verify parent directories exist before creating/editing files.\n\n### Output Handling\n* Stdout and stderr are combined and returned as a string. Output may be truncated if too long.\n* Exit codes are provided in system tags for failed commands.\n* Timeout messages are returned if commands exceed the timeout limit.",
  "parameters": {
    "type": "object",
    "properties": {
      "command": {
        "type": "string",
        "description": "The bash command to execute. You can only execute one bash command at a time. If you need to run multiple commands sequentially, you can use `&&` or `;` to chain them together."
      },
      "timeout": {
        "type": "integer",
        "description": "Optional timeout in seconds for the command execution. If the command takes longer than this, it will be terminated.",
        "default": 120,
        "minimum": 1,
        "maximum": 300
      }
    },
    "required": ["command"]
  }
}
\end{lstlisting}
\end{figure*}

The \texttt{str\_replace\_editor} tool in the baseline follows the standard Anthropic implementation for file viewing and editing via string matching. We omit its full schema for brevity as it follows the standard specification in \citet{schluntz2025swebench}.

\subsection{SWE-Edit Agent Scaffolding}
\label{app:swe_edit_scaffold}

SWE-Edit replaces the \texttt{str\_replace\_editor} with a unified \texttt{llm\_editor} tool that internally delegates to viewer and editor subagents. The \texttt{execute\_bash} tool remains unchanged.

\paragraph{Tool Definition.}
The \texttt{llm\_editor} tool exposes a unified interface for file viewing, creation, and editing, with AI-powered processing for the \texttt{view} and \texttt{edit} commands (Figure~\ref{lst:llm_editor_tool}).

\begin{figure*}[t]
\begin{lstlisting}[language=json, caption={LLM editor tool schema.}, label={lst:llm_editor_tool}, basicstyle=\ttfamily\scriptsize, breaklines=true, breakatwhitespace=false, postbreak=\mbox{\textcolor{red}{$\hookrightarrow$}\space}]
{
  "type": "function",
  "name": "llm_editor",
  "description": "Custom editing tool for viewing, creating and editing files\n* State is persistent across command calls and discussions with the user\n* If `path` is a directory, `view` lists non-hidden files and directories up to 2 levels deep\n* If `path` is a file, `view` uses AI to find and display only the sections relevant to your `query`\n* The `create` command cannot be used if the specified `path` already exists as a file\n\nNotes for using the `view` command:\n* Provide a `query` describing what you're looking for (e.g., \"Where is user authentication handled?\", \"Show me the class definition for User\")\n* The tool reads the file and uses AI to identify relevant line ranges, then displays those sections with line numbers\n* Multiple relevant sections are shown with `... (N lines omitted) ...` separators between them\n\nNotes for using the `edit` command:\n* Provide a clear `instruction` describing what to change and where (identify by function/class/method name)\n* The tool reads the file internally and applies your instruction using AI-powered search-replace\n* Be specific: \"In `MyClass.my_method`, change X to Y\" is better than \"fix the bug\"\n* After editing, the output shows the modified regions",
  "parameters": {
    "type": "object",
    "properties": {
      "command": {
        "type": "string",
        "enum": ["view", "create", "edit"],
        "description": "The command to run. Allowed options are: `view`, `create`, `edit`."
      },
      "path": {
        "type": "string",
        "description": "Absolute path to file or directory, e.g. `/workspace/file.py` or `/workspace`."
      },
      "query": {
        "type": ["string", "null"],
        "default": null,
        "description": "Required for `view` command when `path` points to a file. A natural language query describing what you're looking for in the file. An LLM will analyze the file and return only the line ranges relevant to your query. Examples: 'Where is the authentication logic?', 'Show me the class definition for User', 'Find all functions that handle HTTP requests'."
      },
      "instruction": {
        "type": ["string", "null"],
        "default": null,
        "description": "Required for `edit` command. Detailed instruction describing how to modify the file. Be specific about what changes to make and where (function/class/method name)."
      },
      "file_text": {
        "type": ["string", "null"],
        "default": null,
        "description": "Required for `create` command. The content of the file to be created."
      }
    },
    "required": ["command", "path"]
  }
}
\end{lstlisting}
\end{figure*}

\paragraph{Viewer Subagent Prompt.}
When the \texttt{view} command is invoked on a file, the viewer subagent receives the file contents with line numbers and the user's query. It returns relevant line ranges as a JSON array. The complete system prompt is shown below.

\begin{tcolorbox}[colback=gray!5, colframe=gray!75, title=Viewer Subagent System Prompt, breakable, fontupper=\scriptsize]
\begin{verbatim}
You are an expert code analyzer. Your task
is to identify line ranges in a file that
are relevant to a given query.

You will be given:
1. A file with numbered lines in the format:
   LINE_NUMBER\tLINE_CONTENT
2. A query describing what the user is
   looking for

Your job is to analyze the file and return
the line ranges that are most relevant to
the query. Consider:
- Function/method definitions that match
  the query
- Class definitions related to the query
- Variable declarations or assignments
  relevant to the query
- Import statements if they're relevant
- Comments that explain relevant code
- Any code blocks that implement
  functionality related to the query

OUTPUT FORMAT:
You must output your response as a JSON
array of line ranges. Each range is an
array of two integers [start_line, end_line]
(inclusive, 1-indexed).

Example output:
[[10, 25], [45, 60], [100, 115]]

RULES:
1. Only output the JSON array, no
   additional explanation or comments
2. Line numbers are 1-indexed (first line
   is line 1)
3. Each range should include complete
   logical blocks (don't cut functions/
   classes in the middle)
4. Include a few lines of context before
   and after each relevant section when
   appropriate
5. If nothing in the file is relevant to
   the query, return an empty array: []
6. Ranges should be sorted by start line
   number
7. Merge overlapping or adjacent ranges
8. Keep ranges focused - don't include
   entire files unless the query asks for
   everything

Example 1 - Finding a specific function:
Query: "Where is the calculate_total
function defined?"
Output: [[15, 28]]

Example 2 - Finding multiple related
sections:
Query: "How is user authentication
handled?"
Output: [[5, 8], [23, 45], [102, 130]]

Example 3 - Nothing relevant found:
Query: "Where is the database connection
configured?"
Output: []

Now, analyze the file content and query
provided, and output the relevant line
ranges as a JSON array.
\end{verbatim}
\end{tcolorbox}

\paragraph{Editor Subagent Prompt.}
When the \texttt{edit} command is invoked, the editor subagent receives the file contents and the edit instruction. It outputs modifications in search-replace format or rewrites the entire file when appropriate. The complete system prompt is shown below.

\begin{tcolorbox}[colback=gray!5, colframe=gray!75, title=Editor Subagent System Prompt, breakable, fontupper=\scriptsize]
\begin{verbatim}
You are an expert code editor. Your task
is to analyze a file and make
modifications according to the provided
instructions.

You must output your changes using the
search-replace format shown below. You
can make multiple edits by including
multiple search-replace blocks.

Format for each edit:
<<<<<<< SEARCH
exact lines from the original file to find
=======
new lines to replace them with
>>>>>>> REPLACE

IMPORTANT RULES:
1. The SEARCH block must match the
   original file content EXACTLY,
   including whitespace and indentation
2. You can make multiple edits by
   including multiple search-replace
   blocks
3. If the SEARCH block is empty (no
   content between <<<<<<< SEARCH and
   =======), it means you want to REWRITE
   THE ENTIRE FILE with the content in
   the REPLACE block
4. Each SEARCH block must be unique in
   the file - if there are multiple
   matches, include more context
5. Only output the search-replace blocks,
   no additional explanation or comments

Example 1 - Modifying specific lines:
<<<<<<< SEARCH
def calculate_total(items):
    return sum(items)
=======
def calculate_total(items):
    if not items:
        return 0
    return sum(items)
>>>>>>> REPLACE

Example 2 - Multiple edits:
<<<<<<< SEARCH
import os
=======
import os
import sys
>>>>>>> REPLACE

<<<<<<< SEARCH
def main():
    pass
=======
def main():
    print("Hello, World!")
>>>>>>> REPLACE

Example 3 - Rewriting entire file
(empty SEARCH block):
<<<<<<< SEARCH
=======
#!/usr/bin/env python3
# New file content here
def new_function():
    pass
>>>>>>> REPLACE

Now, analyze the file content and
instruction provided, and output the
necessary search-replace blocks.
\end{verbatim}
\end{tcolorbox}

\subsection{System Prompt}
\label{app:system_prompt}

Both baseline and SWE-Edit agents receive the same system prompt, which describes the task and provides step-by-step guidance. The template variables \texttt{\{\{ instance.repo\_path \}\}} and \texttt{\{\{ instance.problem\_statement \}\}} are populated for each SWE-Bench instance.

\begin{tcolorbox}[colback=blue!3, colframe=blue!50, title=SWE-Bench System Prompt, breakable, fontupper=\scriptsize]
\begin{verbatim}
<uploaded_files>
{{ instance.repo_path }}
</uploaded_files>
I've uploaded a python code repository in
the directory {{ instance.repo_path }}
(not in /tmp/inputs). Consider the
following issue descriptions:

<issue_description>
{{ instance.problem_statement }}
</issue_description>

Can you help me implement the necessary
changes to the repository so that the
requirements specified in the
<issue_description> are met?
I've already taken care of all changes to
any of the test files described in the
<issue_description>. This means you DON'T
have to modify the testing logic or any
of the tests in any way!
Also the development Python environment
is already set up for you (i.e., all
dependencies already installed), so you
don't need to install other packages.

Your task is to make the minimal changes
to non-test files in the
{{ instance.repo_path }} directory to
ensure the <issue_description> is
satisfied.

Follow these steps to resolve the issue:
1. As a first step, it might be a good
   idea to explore the repo to
   familiarize yourself with its
   structure.
2. Create a script to reproduce the error
   and execute it with
   `python <filename.py>` using the
   execute_bash tool to confirm the error
   - **Important:** If testing a Python
     package, add
     `import sys; sys.path.insert(0,
     '{{ instance.repo_path }}')`
     at the top of your script before
     package imports to ensure you're
     testing the local version, not an
     installed version.
3. Edit the source code of the repo to
   resolve the issue
4. Rerun your reproduce script and
   confirm that the error is fixed!
5. Think about edge cases and make sure
   your fix handles them as well

Your thinking should be thorough and so
it's fine if it's very long.
\end{verbatim}
\end{tcolorbox}

\section{Full Experiment Results}
\label{app:detailed_metrics}

\begin{table*}[ht]
\caption{Detailed Performance Metrics on SWE-Bench Verified. Results are averaged over three independent runs for each configuration. ``Succ.'' denotes the success rate of editor tool calls.}
\label{tab:detailed_results_appendix}
\vspace{0.5em}
\begin{center}
\begin{small}
\footnotesize
\setlength{\tabcolsep}{3pt} 
\begin{adjustbox}{max width=\textwidth}
\begin{tabular}{l|cc|cccc|ccc|ccc}
\toprule
\textbf{Config.} & \textbf{Resolved} & \textbf{Rounds} & \textbf{Agent} & \textbf{Editor} & \textbf{Viewer} & \textbf{Total} & \textbf{Output} & \textbf{Cached} & \textbf{Non-Cached} & \textbf{Viewer} & \textbf{Editor}  & \textbf{Succ.} \\
& \textbf{(\%)} & & \textbf{Cost (\$)} & \textbf{Cost (\$)} & \textbf{Cost (\$)} & \textbf{Cost (\$)} & \textbf{Tokens} & \textbf{Input} & \textbf{Input} & \textbf{Calls} & \textbf{Calls} & \textbf{(\%)} \\
\midrule
Baseline & 69.9 & 24.2 & 243.7 & --- & --- & 243.7 & 9632 & 369.8K & 276.7K & 5.78 & 2.86 & 93.4 \\
\quad + Viewer & 70.3 & 23.4 & 215.8 & --- & 9.18 & 225.0 & 9708 & 312.9K & 237.1K & 4.26 & 2.75 & 94.3 \\
\quad + Editor & 71.3 & 22.7 & 263.0 & 5.28 & --- & 268.3 & 8979 & 317.6K & 318.2K & 7.78 & 2.33 & 96.1 \\
\textbf{SWE-Edit} &  72.0 & 20.6 & 179.6 & 5.38 & 15.14 & 200.1 & 9517 & 304.6K & 181.3K & 7.49 & 2.37 & 96.9 \\
\bottomrule
\end{tabular}
\end{adjustbox}
\end{small}
\end{center}
\vspace{-1em}
\end{table*}

\subsection{Per-Run Variance}
\label{app:per_run_variance}
To complement the averaged numbers in Table~\ref{tab:main_results}, we report per-run mean$\pm$std across the three independent runs of the SWE-Bench Verified evaluation in Table~\ref{tab:main_results_std}. The standard deviations are small relative to the reported gains: for example, the resolve-rate gap between SWE-Edit and the baseline ($72.0$ vs.\ $69.9$) is much larger than either configuration's per-run standard deviation, and SWE-Edit attains identical resolve rate ($\pm 0.0$) across all three runs.

\begin{table}[h]
\caption{Per-run mean$\pm$std on SWE-Bench Verified (500 instances, 3 runs).}
\label{tab:main_results_std}
\vspace{0.5em}
\setlength{\tabcolsep}{3pt}
\begin{center}
\begin{scriptsize}
\begin{tabular}{lccc}
\toprule
\textbf{Configuration} & \textbf{Resolved (\%)} & \textbf{Cost (\$)} & \textbf{Edit Succ. (\%)} \\
\midrule
Baseline      & $69.9 \pm 0.6$ & $243.7 \pm 6.5$ & $93.4 \pm 0.8$ \\
\quad + Viewer & $70.3 \pm 1.6$ & $225.0 \pm 5.6$ & $94.3 \pm 0.3$ \\
\quad + Editor & $71.3 \pm 0.2$ & $268.3 \pm 19.3$ & $96.1 \pm 0.2$ \\
\textbf{SWE-Edit} & $\mathbf{72.0 \pm 0.0}$ & $\mathbf{200.1 \pm 16.8}$ & $\mathbf{96.9 \pm 0.1}$ \\
\bottomrule
\end{tabular}
\end{scriptsize}
\end{center}
\end{table}

\subsection{Editor Format Selection in Main Experiments}
\label{app:editor_format_selection}
The Editor subagent prompt (Appendix~\ref{app:swe_edit_scaffold}) supports both \emph{search-replace} edits and \emph{whole-file rewrite} (the latter triggered by an empty \texttt{SEARCH} block). When GPT-5-mini is used as the Editor in the main results (Tables~\ref{tab:main_results}, \ref{tab:opensource}, \ref{tab:downstream}, \ref{tab:editor_scaling}), it selects between these formats per task at inference time. The GRPO-trained Qwen3-8B editor (\S\ref{sec:model_results}) implements the same adaptive selection, learned via reinforcement learning rather than via prompting.

\section{PR-Edit Benchmark}
\label{app:edit_bench}

This section provides implementation details for the PR-Edit Benchmark, including the normalization function used for computing the normalized match reward, the prompt used for GPT-4.1-based equivalence grading, and an example from the dataset.

\subsection{Code Normalization}
\label{app:normalization}
\label{app:canonicalization}

The normalized match reward compares model output against ground truth after canonicalizing whitespace and removing comments. Listing~\ref{lst:normalize} shows the complete implementation. This canonicalization provides a fast, execution-free proxy for edit correctness during training.

\begin{figure*}[t]
\begin{lstlisting}[language=Python, caption={Code normalization function for computing normalized match reward.}, label={lst:normalize}, basicstyle=\ttfamily\scriptsize, breaklines=true, breakatwhitespace=false, showstringspaces=false, postbreak=\mbox{\textcolor{red}{$\hookrightarrow$}\space}]
def normalize_code(code: str) -> str:
    """
    Normalize code by removing comments and normalizing whitespace.
    This allows for comparison that tolerates comment and whitespace differences.

    Note: This uses regex-based heuristics and may incorrectly handle
    comment-like patterns inside string literals (e.g., "http://url" or "# not a comment").
    For most code comparison tasks, this is an acceptable trade-off.
    """
    # Remove multi-line comments first (before single-line to handle edge cases properly)
    # C-style /* */ comments
    code = re.sub(r"/\*.*?\*/", "", code, flags=re.DOTALL)
    # Python docstrings / multi-line strings used as comments
    code = re.sub(r'""".*?"""', "", code, flags=re.DOTALL)
    code = re.sub(r"'''.*?'''", "", code, flags=re.DOTALL)
    # HTML/XML comments
    code = re.sub(r"<!--.*?-->", "", code, flags=re.DOTALL)

    # Remove single-line comments (// for C-like languages, # for Python/shell/etc.)
    code = re.sub(r"//.*$", "", code, flags=re.MULTILINE)
    code = re.sub(r"#.*$", "", code, flags=re.MULTILINE)

    # Normalize all whitespace (spaces, tabs, newlines) to single space
    # This collapses the code into a single line, ignoring all formatting differences
    code = re.sub(r"\s+", " ", code)

    # Strip leading/trailing whitespace
    code = code.strip()

    return code
\end{lstlisting}
\end{figure*}

\subsection{GPT-4.1 Equivalence Grading}
\label{app:gpt_grader}

For the GPT Grader metric in Table~\ref{tab:pr_benchmark}, we prompt GPT-4.1 to assess whether the model's edit is functionally equivalent to the ground truth. The grader receives the original code and two diffs (model output and ground truth), then determines logical equivalence while ignoring cosmetic differences such as formatting, comments, or variable naming.

\begin{tcolorbox}[colback=gray!5, colframe=gray!75, title=GPT-4.1 Grader System Prompt, breakable, fontupper=\scriptsize]
\begin{verbatim}
You are a code analysis expert
specializing in logical equivalence
comparison. You are given the original
code and two diffs showing modifications
to that same original code. Your task is
to determine if these two modifications
are functionally equivalent.

IMPORTANT:
- Focus on logical equivalence, not
  textual similarity
- Consider that different implementations
  can be functionally equivalent
- Ignore cosmetic differences like
  formatting, comments, or variable
  naming

Please analyze these diffs step by step,
then provide your final answer.

REQUIRED OUTPUT FORMAT:
Analysis: [Your detailed analysis here]
Result: [EQUIVALENT/NOT_EQUIVALENT]
\end{verbatim}
\end{tcolorbox}

\begin{tcolorbox}[colback=gray!5, colframe=gray!75, title=GPT-4.1 Grader User Prompt Template, breakable, fontupper=\scriptsize]
\begin{verbatim}
Compare these two code modifications for
logical equivalence:

ORIGINAL CODE:
{original_code}

DIFF 1:
{diff1}

DIFF 2:
{diff2}

Are these modifications functionally
equivalent?
\end{verbatim}
\end{tcolorbox}

\subsection{Dataset Example}
\label{app:dataset_example}

Each instance in the PR-Edit Benchmark consists of three components: (1) the original file content before the pull request, (2) the ground truth file content after merging, and (3) a natural language edit query describing the required modification. Below is a representative example.

\begin{tcolorbox}[colback=gray!5, colframe=gray!75, title=Edit Query, breakable, fontupper=\scriptsize]
\begin{verbatim}
Replace the usage of `assert_image_equal`
with `assert_image_equal_tofile`, and
update imports accordingly to improve the
way image comparisons are being handled.
\end{verbatim}
\end{tcolorbox}

\begin{tcolorbox}[colback=brown!5, colframe=brown!40, title=Original Code (excerpt), breakable, fontupper=\scriptsize]
\begin{verbatim}
import tempfile
from io import BytesIO

import pytest

from PIL import Image, ImageSequence,
    SpiderImagePlugin

from .helper import assert_image_equal,
    hopper, is_pypy

TEST_FILE = "Tests/images/hopper.spider"

...

# for issue #4093
def test_odd_size():
    data = BytesIO()
    width = 100
    im = Image.new("F", (width, 64))
    im.save(data, format="SPIDER")

    data.seek(0)
    with Image.open(data) as im2:
        assert_image_equal(im, im2)
\end{verbatim}
\end{tcolorbox}

\begin{tcolorbox}[colback=teal!5, colframe=teal!40, title=Ground Truth (excerpt), breakable, fontupper=\scriptsize]
\begin{verbatim}
import tempfile
from io import BytesIO

import pytest

from PIL import Image, ImageSequence,
    SpiderImagePlugin

from .helper import
    assert_image_equal_tofile, hopper,
    is_pypy

TEST_FILE = "Tests/images/hopper.spider"

...

# for issue #4093
def test_odd_size():
    data = BytesIO()
    width = 100
    im = Image.new("F", (width, 64))
    im.save(data, format="SPIDER")

    data.seek(0)
    with Image.open(data) as im2:
        assert_image_equal_tofile(im, im2)
\end{verbatim}
\end{tcolorbox}

In this example, the edit requires two coordinated changes: updating the import statement and modifying the function call in \texttt{test\_odd\_size} (final line). For display in the two-column appendix, long import lines in the excerpts above are shown with manual line breaks; in the actual dataset files, each import is a single physical line.

\section{Open-Source Model Evaluation Details}
\label{app:opensource}

\subsection{Model Selection}
Our main experiments use GPT-5, a proprietary model. To verify that SWE-Edit generalizes across model families, we evaluate on three recent open-source reasoning models: Kimi-K2-Thinking~\citep{moonshot2025kimik2thinking}, MiniMax-M2.1~\citep{minimax2025m21}, and GLM-4.7~\citep{zhipuai2025glm47}. These models were selected for two reasons: (1) they represent the latest generation of open-source models with strong reasoning capabilities, and (2) they have undergone substantial agentic training, making them suitable candidates for the challenging software engineering tasks.

\subsection{Inference Configuration}

All three models are configured with \emph{Interleaved Thinking} and \emph{Preserved Thinking} enabled. Interleaved Thinking allows the model to reason before every response and tool call, improving instruction following and generation quality. Preserved Thinking automatically retains reasoning blocks across multi-turn conversations, reusing existing reasoning rather than re-deriving from scratch—reducing information loss and improving consistency for long-horizon agentic tasks.

We use Fireworks AI\footnote{\url{https://fireworks.ai}} for model inference, following each model's official inference settings on SWE-Bench Verified. Common hyperparameters across all models include maximum new tokens of 16,384 and top-$p$ of 1.0. Model-specific exceptions are noted in Table~\ref{tab:inference_config}.

\begin{table}[h]
\caption{Inference hyperparameters for open-source model evaluation. We follow official settings where available.}
\label{tab:inference_config}
\vspace{0.5em}
\setlength{\tabcolsep}{3pt}
\begin{center}
\begin{footnotesize}
\begin{tabular}{lccc}
\toprule
\textbf{Model} & \textbf{Temp.} & \textbf{Top-$p$} & \textbf{Max Tok.} \\
\midrule
Kimi-K2-Thinking & 1.0 & 1.00 & 16{,}384 \\
MiniMax-M2.1 & 1.0 & 0.95 & 16{,}384 \\
GLM-4.7 & 0.7 & 1.00 & 16{,}384 \\
\bottomrule
\end{tabular}
\end{footnotesize}
\end{center}
\end{table}

\subsection{Evaluation Protocol}
Due to computational constraints, we evaluate on the first 100 instances of SWE-Bench Verified rather than the full 500-instance benchmark. Each configuration (baseline and SWE-Edit) is run twice to reduce variance. We use the same baseline agent scaffolding and SWE-Edit architecture as described in Appendix~\ref{app:implementation}, with only the main agent model swapped.

\section{Additional Ablation Results}
\label{app:additional_ablations}

\subsection{Viewer vs.\ Retrieval Baselines}
\label{app:viewer_retrieval}

A natural alternative to an LLM-based viewer is classical code retrieval. We conduct a controlled comparison on 50 held-out PR-Edit instances against (a) dense retrieval using \texttt{text-embedding-3-small} with 30-line chunks and top-3 selection, and (b) BM25 (Okapi) with the same chunking. Ground-truth relevant lines are taken from the PR diff. As shown in Table~\ref{tab:viewer_retrieval}, the LLM viewer attains the highest recall (93.8\%) and the highest F1 (0.272), while still reducing context by 60.3\%. BM25 achieves higher absolute context reduction (64.4\%) but at the cost of severely degraded recall (53.7\%), since natural-language edit queries share few lexical tokens with target code. Dense retrieval improves recall via semantic embeddings but still falls short on both recall and context efficiency. The advantage is most pronounced on long files ($>$300 lines), where the viewer identifies non-contiguous, semantically complete code blocks that fixed-window chunking cannot capture. Note that precision is conservatively defined here: the viewer prompt explicitly returns surrounding logical context (Appendix~\ref{app:swe_edit_scaffold}, Rules~3--4), not only the exact diff lines.

\begin{table}[h]
\caption{Viewer vs.\ retrieval baselines on 50 held-out PR-Edit instances. ``Ctx.\ Red.'' is the percentage of the input file omitted from the returned snippets.}
\label{tab:viewer_retrieval}
\setlength{\tabcolsep}{5pt}
\begin{center}
\begin{scriptsize}
\begin{tabular}{lcccc}
\toprule
\textbf{Method} & \textbf{Recall} & \textbf{Prec.} & \textbf{F1} & \textbf{Ctx.\,Red.} \\
\midrule
LLM Viewer (GPT-5-mini)        & \textbf{0.938} & \textbf{0.179} & \textbf{0.272} & 60.3\% \\
Dense (text-embedding-3-small) & 0.868 & 0.092 & 0.140 & 28.6\% \\
BM25 (Okapi)                   & 0.537 & 0.056 & 0.083 & \textbf{64.4\%} \\
\bottomrule
\end{tabular}
\end{scriptsize}
\end{center}
\end{table}

\subsection{Training Dynamics}
\label{app:training_dynamics}

\begin{figure}[h]
\centering
\includegraphics[width=\columnwidth]{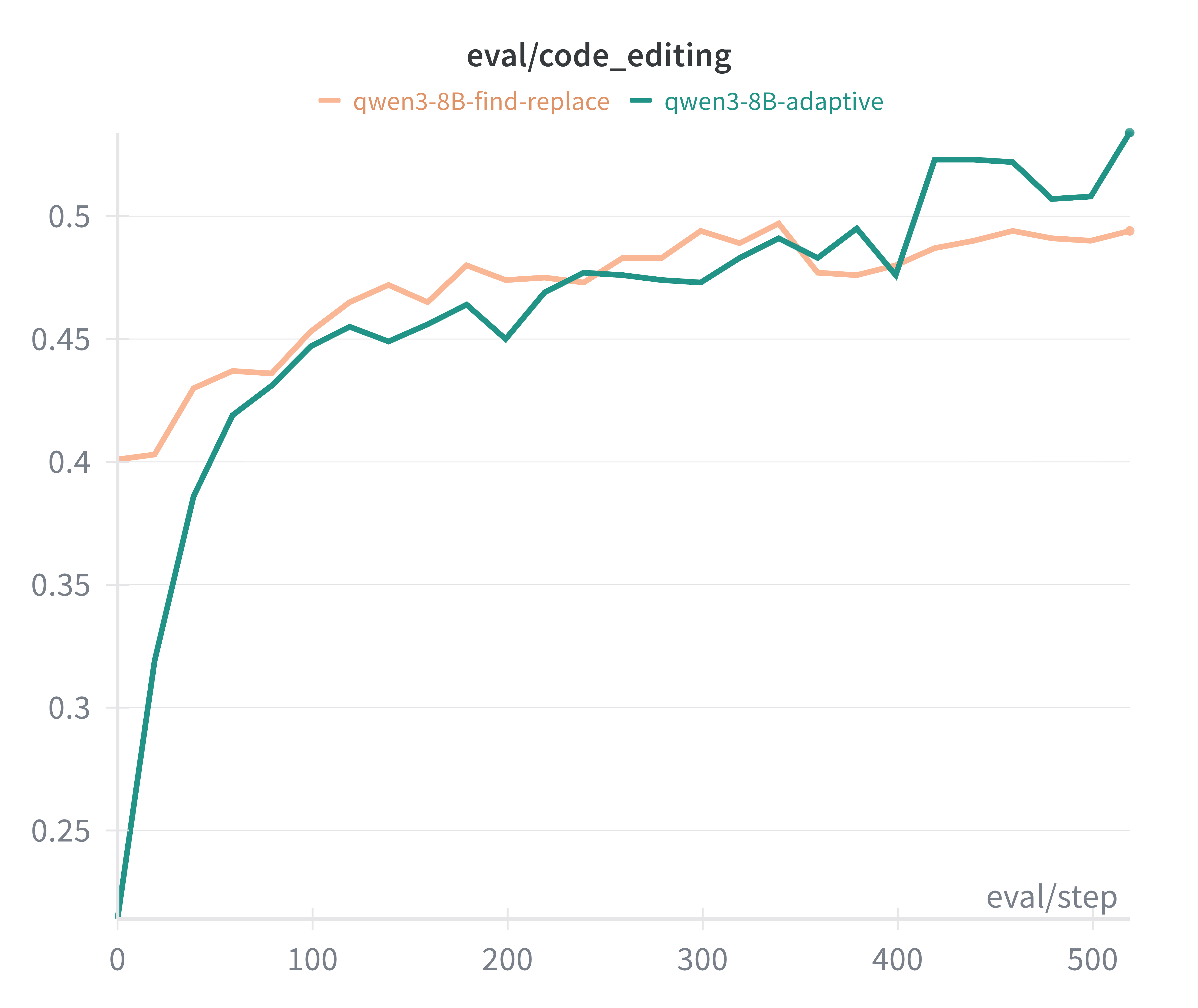}
\caption{Training dynamics for fixed vs.\ adaptive format selection. The y-axis is validation reward (normalized match) and the x-axis is the rollout step. While fixed find-replace starts higher (simpler format, easier to learn), adaptive training surpasses it by learning when to invoke whole-file rewrite.}
\label{fig:training_dynamics}
\end{figure}

Figure~\ref{fig:training_dynamics} illustrates the training dynamics for fixed vs.\ adaptive format selection. The fixed find-replace policy starts with higher validation reward---unsurprising, as find-replace is simpler and most training examples involve localized edits. However, adaptive training converges to a higher final reward, as the model learns to invoke whole-file rewrite for the subset of tasks where find-replace struggles. This confirms our hypothesis: a single format cannot optimally serve all edit types, and learning to select adaptively yields consistent gains.

\subsection{Scaling the Editor Model}
\label{app:editor_scaling}

We examine whether stronger models yield proportional gains in the editor role by replacing GPT-5-mini with GPT-5. As shown in Table~\ref{tab:editor_scaling}, GPT-5 improves resolve rate by only 0.4\,pp (72.0\% $\rightarrow$ 72.4\%) while increasing editor cost by 5.8$\times$ (\$5.4 $\rightarrow$ \$31.2 per run). The marginal gain suggests that scaling alone is not cost-effective. Compared to Table~\ref{tab:downstream}, GRPO-based adaptive editor training yields a 12.5\,pp increase in edit success on Qwen3-8B---an order of magnitude larger than the +0.6\,pp obtained by scaling from GPT-5-mini to GPT-5---indicating that format-level decision-making is a learnable skill rather than a capability that emerges only at scale.

\begin{table}[h]
\caption{Effect of editor model scale. GPT-5 provides minimal accuracy gain at 5.8$\times$ the cost.}
\label{tab:editor_scaling}
\setlength{\tabcolsep}{5pt}
\begin{center}
\begin{scriptsize}
\begin{tabular}{lccc}
\toprule
\textbf{Editor Model} & \textbf{Resolved (\%)} & \textbf{Edit Succ.\,(\%)} & \textbf{Cost (\$)} \\
\midrule
GPT-5-mini & 72.0 & 96.9 & 5.4 \\
GPT-5 & 72.4\,{\scriptsize(+0.4)} & 97.5\,{\scriptsize(+0.6)} & 31.2\,{\scriptsize(5.8$\times$)} \\
\bottomrule
\end{tabular}
\end{scriptsize}
\end{center}
\end{table}

\subsection{Scaling the Viewer Model}
\label{app:viewer_scaling}
We check whether a stronger viewer would help. Replacing GPT-5-mini with GPT-5 in the viewer role yields SWE-Bench resolve rate of 70.2\% vs.\ 70.3\% at matched total cost (\$225)---within run-to-run variance (Table~\ref{tab:main_results_std})---indicating that the viewer role is well-served by a smaller, cost-efficient model.
\end{document}